# Web search engine based on DNS[1]

## Wang Liang[1], Guo Yi-Ping[2], Fang Ming[3]

[1, 3] (Department of Control Science and Control Engineering, Huazhong University of Science and Technology, WuHan, 430074 P.R.China)

[2] (Library of Huazhong University of Science and Technology, WuHan 430074 P.R.China)

**Abstract** Now no web search engine can cover more than 60 percent of all the pages on Internet. The update interval of most pages database is almost one month. This condition hasn't changed for many years. Converge and recency problems have become the bottleneck problem of current web search engine. To solve these problems, a new system, search engine based on DNS is proposed in this paper. This system adopts the hierarchical distributed architecture like DNS, which is different from any current commercial search engine. In theory, this system can cover all the web pages on Internet. Its update interval could even be one day. The original idea, detailed content and implementation of this system all are introduced in this paper.

## 1. Introduction

Because the WWW is a large distributed dynamic system, current search engine can't continue to index close to the entire Web as it grows and changes. According to the statistical data in 1998[1], the update interval of most pages database is almost one month and no search engine can cover more than 50 percent of pages on Internet. Till now, these data are still available. Converge and recency problems become the bottleneck problem of current web search system.

Despite the serious problems of web search engine design technology, very little academic research has been done on them in these years. Commercial profit also make many search engine researches keep secret. Most efforts to solve these bottleneck problems all adopt the distributed architecture, but not the centralized architecture of current systems. In fact, one of the first search engines, Harvest [2] is just a representative distributed web search system. Latterly, based on Harvest, Cooperate Search Engine (CSE) [3] is developed. These two methods require each web site to index their web documents and provide interface for search engines. These approaches can reduce the update interval and network traffic, but none of them are widely implemented. This is mainly because not all administrators of web sites agree to index their pages for search engines, its search speed can't be ensured either. Reference [4] gave a practical idea to build a distributed search engine. In this paper, Author advised to share the databases of the search engine and introduced a layered architecture to improve the access to data on the Internet. But in this paper, the author didn't give a practical method on how to implement his idea. We develop our new search engine based on these previous works.

## 2. Basic idea of web search engine based on DNS

### 2.1. Origin

The original idea of new system could be found in the history of WWW. When the DNS comes into being, there are only hundreds of web sites, so we can put DNS table in single server. When the web sites reached million level and scattered in different places, several DNS can't work efficiently. So DNS developed into a distributed hierarchical system. Now almost all the universities and big organizations have their own DNS server in local network. All the web sites on Internet are efficiently managed in this system.

But DNS only provides navigation service. We also need searching the WWW. So web search engines appeared. First Yahoo, then Google. For some reasons, all commercial search engines adopt centralized architecture. So with the rapid increase of WWW, they also meet the problems that primal DNS meet. There have been billions of pages scattering all over the world. Current search engine have to download them to a centralized database system again

---

[1] Corresponding author
Wang Liang, E-mail: guoypm@hust.edu.cn, Phone: +86-27-87542739. fax: +86-27-87554415

and again. The coverage and update interval can't be ensured in this system. Obviously centralized architecture is not appropriate to manage the distributed WWW. Unsuitable architecture is the key for most bottleneck problems in current search engine. So a better solution must choose a completely different architecture.

Adopting the experience of DNS may be a good selection. The distributed hierarchical architecture of DNS may also provide an efficient architecture to build the new web search engine. Furthermore, Since DNS can index the name of each site, could they index all the web pages in different sites? So idea of "search engine + DNS" appeared.

*2.2. Basic Architecture*

As the basic idea of new search engine based on DNS, its architecture is as same as DNS, which is shown in fig1.

There are three layers in this system. The third layer is the third level Domain, always corresponding to some organizations like a university. The second layer is the second level Domain, the sub net of a country. The top layer corresponds to each country. Its three layers strictly correspond to different levels of DNS.

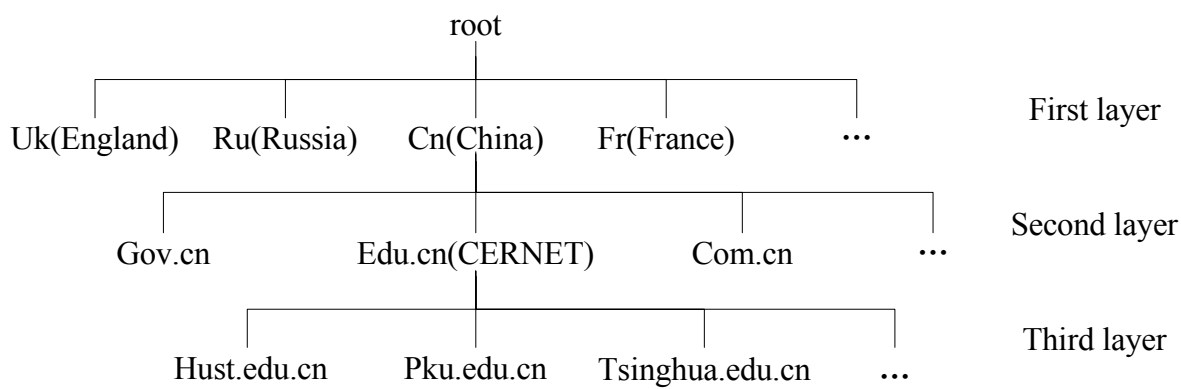

Fig.1. Architecture of web pages search engine based on DNS

In this architecture, we can simply download all the web pages in the bottom layer and send them to the servers in the higher layer. Because all the pages extracting work is done by the different servers in bottom layer, which always corresponds to local network, the update interval of whole system could be reduced to a reasonable level. So the recency problem could be solved. But in this method, the databases in the top layer will be fairly large. We may have to adopt distributed computing or other complex technologies to design such system. All in all, building a database system that can mirror the whole Internet is almost an impossible mission. We have to find a practical method to realize the basic idea of this system.

## 3. Technology related to search engine

In this section, we will give a high overview of some technologies related to search engine. First, we will give a brief introduction for basic search engine technology. Then as the key technology of search engine, two kinds of ranking algorithms are presented. Web search engine is just one kind of information retrieval system. Three different information retrieval systems are introduced at last. All these technologies will be employed in the design of new system.

*3.1. Basic search engine technologies*

Most practical and commercially operated Internet search engines are based on a centralized architecture that relies on a set of key components, namely Crawler, Indexer and Searcher, whose definitions are introduced as follows[5].

- Crawler. A crawler is a module aggregating data from the World Wide Web in order to make them searchable. Several heuristics and algorithms exists for crawling, most of them are based upon following links.
- Indexer. A module that takes a collection of documents or data and builds a searchable index from them. Common practices are inverted files, vector spaces, suffix structures and hybrids of these.
- Searcher. The searcher is working on the output files from the indexer. The searcher accepts user queries,

runs them over the index, and returns computed search results to issuer.

*3.2. Two kinds of ranking technologies*

Now there are two main page ranking technologies.

- Text information retrieval technology [6]. Web search technology is derived from the text information retrieval technology, which employs the so called TF*IDF search and ranking algorithm which assigns relevance scores to documents (the term *documents* here refers to text files in general, which also applies to WWW pages).The big difference between them is that the former can make use of rich tags in the web documents, which gives some hint about the documents. According to a specific strategy, different tags are given different weights so that the search quality can be improved. This is a basic technology and used by many early search engines such as Excite, Infoseek, etc.
- Hyperlink analyzing technology [7]. Its basic idea is that the ranking score of a page is determined by the number of links in other sites that point to this page. There are two representative algorithms for this technology. Based on the simulation of the behaviors of uses when they surf on the web, Larry page brings the model called PageRank [8], which presents an algorithm to calculate every web page's weight. In IBM research lab's "clever" system, a group demonstrates an HITS model [9], which automatically locates two types of pages: authorities and hubs. The first type provides the best sources of information on a given topic, and the second type points to many valuable authorities. This technology gives higher weight to the pages many other pages link to. The newly published pages on web will be ignored.

If we design a search engine in a small scope of Internet like in a university, text information retrieval technology will be enough. Hyperlink analyzing technology is more effective in large scale like in a country

*3.3. Three kinds of information retrieval system*

According to the architecture, there have been three kinds of information retrieval systems, which are introduced as follows.

- Centralized search system. This system has its own data collecting mechanism, and all the data are stored and indexed in a conventional database system. Although many web search engines download web pages and provide service by thousands of servers, they all belong to this kind according to their basic architecture.
- Metadata harvest search system. Normally Metadata is much smaller than data itself. So when we can't store all the data in a database system or need to integrate different kinds of information resource like video, PDF, web pages in a system, we can harvest the metadata from different sub databases and build a union metadata database. This database can provide the search service just like the conventional database system. Some library database systems based on OAI [10] just adopt this method.
- Distributed search system. When the data source is large enough that even the metadata can't be efficiently managed in a database system, we can choose distributed system.Distributed information retrieval system has no its own actual record database. It just indexes the interface of sub database system. When receiving a query from a user, main system will instantly obtain the records from sub databases through their search interfaces. The limitation of this system is that the number of sub databases can't be many, otherwise the search speed can't be ensured. A famous system is InfoBus system in Stanford digital library project [11].

There are two main factors to determine the architecture of an information retrieval system, the size and diversity of data source. Normally, with the increase of size and diversity of data source, we can select centralized system, metadata harvest system and distributed search system respectively.

## 4. Realization of web search engine based on DNS

According to the different characters of three layers, we adopt the different architectures and ranking algorithms to build the new system. Three layer's search systems are introduced as follows.

*4.1 Third layer: centralized search system*

The node of this layer always corresponds to an organization. Normally, a centralized search system can efficiently manage all its web pages. The node in this layer is just a conventional web search engine. Only difference is that its search scope is limited in a third level Domain like a university. This search system comprise of three parts, crawler, indexer and searcher, which are introduced respectively.

*4.1.1. Crawler*

Crawler of this layer will download all the pages in a third level Domain. For example, "hust.edu.cn" is the domain name of our university, so the server under the domain of "hust.edu.cn" like the department of Computer Science's server "cs.hust.edu.cn" can be easily found in the DNS server of our university. By this means, the crawler can download all the web sites' pages in this Domain.

The work of crawler is arranged by single web site. When a spider visits a web server, it will download all the pages in this site. It will stop working when it encounters a URL linking to other site. This kind of URL is called "stop URL" .The content in this "stop URL" is also treated as valuable matter of this site and is downloaded too. This theory has some differences with conventional search engine, whose crawler will freely surf in WWW and have no stop URL. We have to design complex algorithms to direct crawler. But in new system, its crawler just need download all the pages in each site and not consider the intricate relation between different sites.

*4.1.2. Indexer*

Normally, the key issue in indexer is the appropriate selection of metadata. In text information retrieval technology, every word of document is indexed with its ranking score. We also use this method to index the web pages. By this means, the rank score of keyword is determined by its position and frequency information. The other information such as tags, encoding and abstract will also be used to describe a web page. Moreover, we can also use W3C's Ontology [12] model or other advanced technologies to index the web pages.

*4.1.3. Searcher*

Providing user interface and processing the search results are main functions of searcher. How to rank the pages is the key issue. In this layer, we use the text Information retrieval technology to rank the results. This is because its web pages are limited in a small area like a university, but the Hyperlink Analysis is more useful in large scale. This paper is mainly concerned with how to build the search system. Detailed ranking algorithm will be determined in its standard protocols.

*4.2. Second layer: metadata harvest search system*

This layer will provide the search service that covers all the web sites under a second level Domain. Metadata harvest system is adopted in this layer. A third layer node like "hust.edu.cn" corresponds to our universities. Most universities have no more than 100 thousand pages, so a centralized search system can work efficiently. But a second layer node "edn.cn" includes the web pages of all the universities in China. Centralized search system may not ensure the recency and coverage of its database. So we use metadata harvest system in this layer. The search engine of this layer has only two parts, indexer and searcher.

*4.2.1. Indexer*

Its data is downloaded from the databases in the corresponding third layer's nodes. For example, the search engine corresponding to the domain "edn.cn" will obtain its data from the search engine's databases in thousands of universities in China, but not directly extract millions of web pages itself. By this means, the update interval could be much shorter than conventional methods. Only storing the metadata will also make its database fairly small. This web metadata harvest protocol refers to the OAI [10] protocol, which is a mature metadata harvest system and has been adopted in many library systems.

*4.2.2. Searcher*

A notable problem that should be concerned with is the overlap of web pages when harvesting the metadata. In the third layer, when the crawler extracts pages from a site, it also gets some pages which don't belong to this site (Stop URL). So when harvesting the metadata, some pages may appear many times. As the download principle in the third layer, the overlap number (no including this page) is just the number of pages of other sites which direct to this page. According to the theory of Hyperlink Analyzing, the ranking score of this page could be calculated by this number. New system employs a simple method to realize the basic idea of Hyperlink Analyzing. Obviously

Hyperlink Analyzing technology will be more efficient to rank the search results in this layer. Detailed content like how two layers cooperate to transfer the metadata and the finial ranking algorithm will be defined in its protocols.

*4.3. Top layer: distributed search system*

Because the number of second layer's node (number of second level's Domain name) is no more than ten, and storing the metadata of billions of web pages in single system is still a big challenge, search engine in this layer could be a distributed information retrieval system. It has merely one part, searcher, no crawler, and no indexer.

There are three issues when designing a distributed search system [13].1 the underlying transport protocol for sending the message (e.g., TCP/IP). 2 a search protocol that specifies the syntax and semantics of message transmitted between clients and servers.3 a mechanism that can combine the results from different servers. These problems are detailed as follows.

1 Communication protocol. In this system, SOAP is adopted as the basic protocol for communication.

2 Search protocol. Its search protocol is based on Webservice. Webservice use the SOAP as its fundamental protocol and provides an efficient distributed architecture. We refer to the SDLIP [14] and Google's search Webservice to define the format of query and results. All the search engines in the second layer should provide the standard search Webservice according to a standard protocol. Search engine in top layer just needs to index all the Webservice in lower layer.

3 Combining the results. In this system, the key problem for the combination of results is page ranking. In the second layer, the ranking score of a page is calculated by its overlap number. In this layer, we also use this theory to rank the pages. The ranking score of same page from different databases will be added up and a finial ranking list of pages will be produced.

The work theory of this search engine is just like that of Meta search engine [15], which has no its own pages database, but an index of the search interface of other engines. Some experience of Meta search engine research like how to combine the search results could be adopted in this search system. Because the sub search engines of this system are strictly arranged and comply with the same protocol, its performance will be much better than any current Meta search engine. The search engine in this layer will provide the search service in the scope of a country, which will cover the most of search request on Internet. If you want to search in several countries, you can use their search Webservice in top layer to build a special search engine.

## 5. Application of web search engine based on DNS

The next question is how to use this system. Because all the nodes of this system are integrated search engines, how to help the user find the appropriate search engine is the key for its application. As an application software system, we normally use class tree of OO (Object Oriented) model to describe this system. We chose a namespace DRIS (Domain resource integration system) for this system, which means integrating the information resources in different Domains. The class tree of search engine based on DNS is shown in Fig 2.

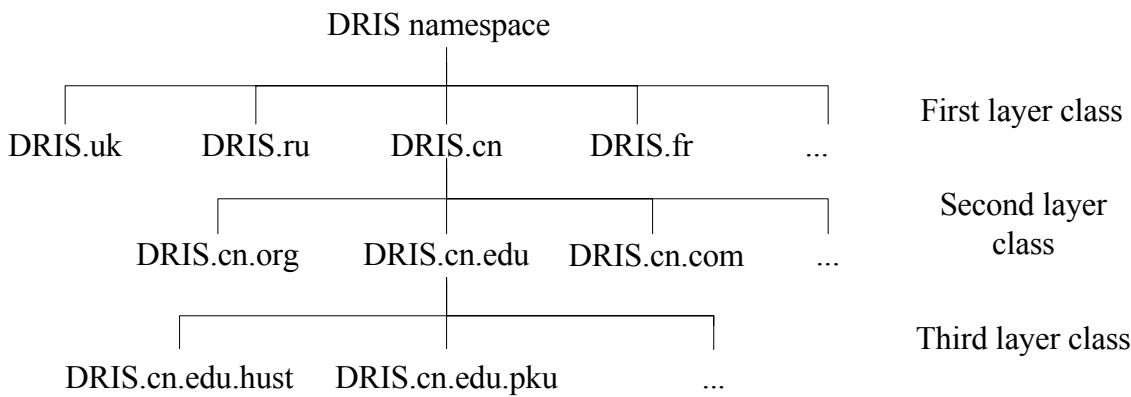

Fig.2. OO model of web search engine based on DNS

All the nodes in this system are arranged under the namespace "DRIS" and can be treated as the child class of DRIS. These child classes and their function are realized in different servers. We define some basic rules for the management and application of this system.

1 All the nodes of DRIS will provide the search service in the form of standard Webservice.

2 All these search Webservice are organized by the inheriting relation, but its realization is different from that of conventional OO model. The node in lower layer can inherit the class in higher layer by citing its Webservice. But because search engine based on DNS is a distributed system, the nodes in higher layer lie in different servers and can't know its child class. So the node in higher layer will have a special mechanism to index the search interface of all its child class.

3 The Webservice provides the service by its URL. Every node of this system is an integrated search engine. So for ender users, how can they find the appropriate search service quickly? For example, I want to find the search Webservice of a university, but what's its URL? So for the convenience of end user, we define a special rule for the location of search engine server. If main class name of a search Webservice is "DRIS.Class name", its Webservice should provide the service through URL of "http://DRIS.Reversed Class name". Because class name of each node corresponds to the each Domain name, this rule can also be described as that all search engines should provide the search service through URL of "DRIS. Domain name" and the main class name of this Webservice will be "DRIS. Reversed domain name". For example, the domain name of our university is "hust.edu.cn". So our search engine server will provide the search service through URL of "http://DRIS.hust.edu.cn" and its main class name is just "DRIS.cn.edu.hust ". So if this system is implemented, all Domain names "DRIS. Domain name" should be left to search engine servers.

This system provides the standard search Webservice in different scope of Internet, which will be the excellent data source for some personal search systems. According to the interests, user behavior and other personal information, personal search system can automatically select the appropriate data source and rearrange the search results to match user's preference. The ranking score calculated by hyperlink analyzing and other technologies is only regarded as the reference score. Personal information will be the most important factors to rank the search results.

There have been some researches for personal web search system. Most of these systems get results by post-processing the results of existing commercial search engines, or produce small scale "individualized" search engines. In [16], analyzer can accumulate, categorize and personalize web usage information, and give a personalized, knowledge-driven search system that helps the user find the informative web site. In search system of [17], feedback information produced by the user's accessing lists can influence the search results and search engine can provide self-adaptability. Web search engine based DNS will be an efficient platform for these researches.

## 6. The advantage of web search engine based on DNS

To make the search results more precise, this search system is completely divided into two independent parts, a web information retrieval infrastructure and the personal search system. To build an efficient information retrieval infrastructure, this system adopts the hierarchical distributed architecture like DNS. Seeing from the top layer, its search service is as same as current commercial search engine, but its three components are in separate places. Its crawler lies in the different local network. Indexer is in the second layer. Searcher provides the search service in the top layer. Three components are connected by metadata harvest and other technologies. All the nodes of search engine based on DNS are integrated search engine and provide search service in different scopes. Normally there are three principles to judge a search engine.

1 The coverage of the search engine. The more pages are indexed, the better a search engine is. Because the work of crawler is arranged by Domain, as long as the sites could be found in DNS, their pages all can be covered in this system. In theory, the search engine based on DNS can cover almost all the pages on the Internet.

2 The update interval of search engine. All the downloading work of this system is carried out by many crawlers in the third layer, which always correspond to local network. Their update works only need a short time. In the second layer, the metadata upload process won't spend much time either. The top layer hasn't the actual record and need not update. So the update interval of this system will be much shorter than current search engine.

3 Precise of search results. Current search engine never consider what's the user's favorite. Everyone gets the same results for a query, which can't express our interests. All the nodes of this system will provide standard search interface. Many intelligent personal search systems could be designed according to your personal information. This system will be their data source. More precise search results could be obtained in these personal search engines.

## 7. Implementation and management of new system

Although search engine based on DNS gives us an excellent and promising solution for the new Internet search system, this can't ensure its establishment. We will meet a common problem: who are willing to build such system. Finding the request of users is the key for any new system. The distinct architecture of search engine based on DNS will ensure it's a practical system. In the bottom layer of this system, it will build the search engine in the scope of an organization like a university. Now some universities have purchased the commercial search engine for their school network. If a free system appears, most organizations will choose it. Once the systems in bottom layer are completed, we will have the foundation for the higher layer system. Solving some urgent problems of their own and then bringing benefit to others may be the real secret for the implementation of this system.

The realization section of this paper just gives a basic principle to construct this search engine. This principle could be adapted for different conditions. For example, in some top level Domain or second level Domain, there may be only small amount of web pages, which can be efficiently managed in a database system. A centralized search system will be enough. But for some big web sites like Microsoft, their data may reach GB level. Just providing a search interface may be more appropriate than being downloaded again and again.

The management of system is as same as DNS. It's built and managed by its users and coordinated by a public organization. In the bottom layer, the organizations can define the metadata upload rules by themselves. This method can avoid copyright and privacy problems in current search engine. Paper [18] has some considerations about the method to control the publication in a distributed web search system.

Now we are building the first test bed on CERNET (China education and research network), which includes all the universities in China. In CERNET, most sites strictly comply with the rules of DNS and are under the Domain of "edu.cn". All the universities are also connected by high speed communication network. It provides an ideal environment for the first experimental system.

But in fact, just in our university, there are few sites which don't use our domain name. They may be in "net" or other Domains. Our search engines still index these sites for the convenience of its users. But we have to design some additional mechanisms in metadata upload process to avoid the confusion in the higher layer. On the other hand, our university also encourages us to use our university's Domain name for our web sites. Moreover, for some reasons, few universities may be not willing to build their search engine servers. For some former distributed search engines, they can't do anything else in this condition. But in this system, their pages could be downloaded in the higher layer without their permission. Although it may be not very reasonable, most current search engines just do like this. These conditions will also happen in other Domains. We may have to find some tradeoff when building this system.

## 8. Conclusions

Search engine based on DNS extend the "navigation" function of DNS to web "search" function. New system has obvious improvement in coverage, recency and precise. The kernel idea of this system is that search should be the internal function of Internet and everyone should have his own search engine. Moreover, new system is a public information retrieval system, but not a commercial search engine. But some companies can design better search system based on this public platform. In fact, almost all the Internet technologies, from TCP/IP to E-mail, are public technologies, but the better commercial service could be designed based on them. It may be a basic principle for the continued development of Internet.

## 9. Future works

1 Semantic web research. HTML is an easy tool to build the web pages, but it also brings many troubles for search engine. So XML is proposed by W3C.This technology can bring the order to the web. But it's a difficult mission to persuade all the web designers to adopt XML for their pages. XML will give many benefits for search engine, but it may be not ideal tools for web pages designer. It's a dilemma for web page designer and search engine designer. Some advanced technologies based on XML such as RDF and Ontology model (Semantic web) [19] are also proposed by W3C, but they all meet this trouble. The web search engine based on DNS may give a promising solution for this dilemma. In the bottom layer of this system, we can use XML based technologies to index the web pages. So for web designer, they can still use HTML or other formats to build the web pages. But for search engine designer, they will get the standard XML data from different servers of web search engine based on DNS. Indexing the web in the format of XML is an easy mission. But how to index the pages by RDF or Ontology model and convert the current WWW to Semantic WWW is still a big challenge.

2 Internet information infrastructure research. Web page is only one kind of information resource on Internet. There are many other kinds of resources such as FTP, Video, many special database, etc. As a common Internet user, you may always complain too many unrelated search results in web search engine, but when you come to library, they may find it's just the beginning. Almost in every university's library, there are nearly one hundred different kinds of digital resources such as IEEE, SDOS, Springer, many self-build databases etc. We can also obtain information from web page search engine and other public information resources. To find the comprehensive and precise information, we have to search in many search interfaces one bye one. We need also be familiar with different query rules of every database. But the Internet information resource is still exploding. Searching the whole Internet is becoming an impossible mission. Everyone wants to get all kinds of information on Internet in one system and never care where the record lies in. It's no longer a problem in digital library, but a serious problem of whole Internet. Just like the web resource, all the information resources on Internet scattered all over the world. Although the "physical Internet" is integrated system, the "information Internet" is still a fragmentized world. The distinct architecture of DNS may also be a good architecture to build an efficient Internet information infrastructure that can connect and cover all the information resources on Internet. As the extension of web pages search system based on DNS, DRIS (Domain Resource Integration System) was proposed for this issue. There have been some discussions [20] in IETF for it.

These works may give some promising research directions for the next generation of search engine technology, which will directly give an answer for your question like who is the creator of Google, but not piles of hyperlinks.

**Acknowledgments** The research described here was conducted as part of HUST (Huazhong University of Science and Technology) Digital Library project, supported by CALIS (China Academic Library & Information System) and national "211" project. IETF also give many suggestions and helps for this research.